\documentclass[doublecol,linenumbers]{epl2}
\usepackage{amssymb}
\usepackage{amsmath}
\usepackage{soul}

\newcommand{\bea}{\begin{eqnarray}}
\newcommand{\eea}{\end{eqnarray}}

\def\beq{\begin{equation}}
\def\eeq{\end{equation}}
\def\etal{{\it et al.}}

\title{Valence fluctuations in a lattice of magnetic molecules: 
application to iron(II) phtalocyanine molecules on Au(111)}
\shorttitle{Valence fluctuations in a lattice of FePc molecules} 

\author{J. Fern\'andez\inst{1}  \and A. A. Aligia\inst{1} \and A. M. Lobos\inst{2},\inst{3}}

\institute{                    
  \inst{1} Centro At{\'o}mico Bariloche and Instituto Balseiro, 
Comisi\'on Nacional de Energía At\'omica, 8400 Bariloche, Argentina\\

\inst{2}  Facultad de Ciencias Exactas Ingenier\'ia y Agrimensura, Universidad
Nacional de Rosario and Instituto de F\'isica Rosario,  Bv. 27 de Febrero
210 bis, 2000 Rosario, Argentina\\
    
  \inst{3} Joint Quantum Institute and Condensed Matter Theory Center, Department of
Physics, University of Maryland, College Park, Maryland 20742, USA
}
 

\abstract{
We study theoretically a square lattice of the organometallic Kondo adsorbate iron(II) 
phtalocyanine (FePc) deposited on top of Au(111), motivated by recent scanning tunneling microscopy experiments. 
We describe the system by means of an effective Hubbard-Anderson model, where each molecule has degenerate 
effective $d-$orbitals with $xz$ and $yz$ symmetry, 
which we solve for arbitrary occupation and arbitrary
on-site repulsion $U$. To that end, we introduce a generalized slave-boson mean-field approximation (SBMFA) 
which correctly describes both the non-interacting limit (NIL) $U=0$ and the strongly-interacting limit 
$U \rightarrow \infty$, where our formalism reproduces the correct value of the Kondo temperature 
for an isolated FePc molecule.
Our results indicate that while the isolated molecule can be described by an SU(4) Anderson model in the Kondo
regime, the case of the square lattice corresponds to the intermediate-valence regime, 
with a total occupation of nearly 
1.65 holes in the FePc molecular orbitals. 
Our results have important implications for the physical interpretation of the experiment.}

\begin{document}

\maketitle

The advances in nanotechnology and the degree of control of different parameters in  systems with
strongly interacting electrons provide a rich and challenging field for the theoretical understanding 
of strongly-correlated materials, and for future applications in spintronics and quantum 
information\cite{Wolf01_Spintronics_review}.
The strong repulsion between $3d$ electrons in the $3d$ series of transition-metal atoms (TMAs) has led 
to the observation of the Kondo effect in several systems in which these atoms 
\cite{Li98_Kondo_effect_on_single_adatoms, Madhavan98_Tunneling_into_single_Kondo_adatom, knorr02} 
or molecules containing them 
\cite{Zhao05_Science, Iancu06_Manipulation_of_Kondo_effect, Gao07_Site_specific_Kondo_effect,Tsukahara10_Evolution_of_Kondo_resonance, Franke11_Competition_of_Kondo_and_SC_in_molecules,Minamitani12_SU4_Kondo_in_FePc_molecules, DiLullo12_Molecular_Kondo_chain} 
are absorbed on noble-metal surfaces.
In its simpler form, the Kondo effect arises when a TMA has a (nearly) integer occupation
of an odd number of electrons and spin 1/2, and this spin is ``screened''
by a cloud of conduction electrons in the metal, building a many-body singlet 
ground state.
As a consequence, a narrow Fano-Kondo antiresonance (FKA) appears in the differential conductance in 
scanning tunneling spectroscopy (STS) experiments. 
More exotic Kondo effects involving even occupation and spin 1 \cite{parksjj,serge},
 and orbital degeneracy \cite{jarillo05, tetta12, Minamitani12_SU4_Kondo_in_FePc_molecules} were also observed. 
In particular, it has been shown that STS for a system consisting on a single
iron(II) phtalocyanine (FePc) molecule (see the inset of Fig. 1) deposited on top of clean Au(111)
(in the most usual ``on-top'' configuration) can be explained in terms of an effective 
SU(4) impurity Anderson model \cite{Minamitani12_SU4_Kondo_in_FePc_molecules}. 
The four degrees of freedom correspond to the intrinsic spin 1/2 degeneracy times 
two-fold degeneracy between molecular orbitals with the same symmetry as the $3d_{xz}$ and
$3d_{yz}$ states of Fe, where $z$ is the direction perpendicular to the surface. 
The Kondo model is the odd-integer valence
limit of the impurity Anderson model, in which localized electrons
(like those of the $3d$ shell of a TMA) are hybridized with a conduction band. 

Experiments in which 
an array of effective ``impurities'' (TMAs or molecules containing TMAs) are deposited
on metal surfaces constitute a further challenge to the theory because they should be described by the 
periodic Anderson model, for which neither exact solutions nor accurate treatments, like the 
numerical renormalization group, exist. Tsukahara {\it et al.} studied the STS of a square lattice 
of FePc molecules on Au(111) \cite{Tsukahara10_Evolution_of_Kondo_resonance}. They observed a splitting of the FKA near the Fermi level. 
A first hint to the correct description of the system came soon when the corresponding experiments for
an isolated FePc molecule could be well described by the effective SU(4) 
impurity Anderson model \cite{Minamitani12_SU4_Kondo_in_FePc_molecules}.
A natural step to construct a lattice model is to introduce an effective hopping between the molecules, 
which leads to a two-band Hubbard model when the hybridization $H_{mix}$ with the metallic states is neglected. 
When $H_{mix}$ is included, one obtains an 
orbitally-degenerate Hubbard-Anderson model.
Lobos {\it et al.} \cite{Lobos14_SU4_Kondo_lattice} derived this model and solved it using 
a slave-boson mean-field approximation (SBMFA) for infinite on-site repulsion $U$ 
\cite{coleman84}. These authors were able to reproduce the observed splitting
of the FKA, which in simple terms can be interpreted as arising from split van Hove singularities due to 
the anisotropic hopping terms in the Hubbard part of the model. 

However, there are several indications 
that the theory needs to be improved in order to reproduce the experimental details. 
For instance,
the energy necessary to add the first hole $E_d \approx -0.1$ eV [see Eq. (1)]
was estimated from \textit{ab-initio} 
calculations in the local density approximation (LDA), which obtained spectral density of Fe $3d_{xz}$ and $3d_{yz}$ 
states 0.1 eV above the Fermi energy $\epsilon_F$, which we set as the origin of energies ($\epsilon_F=0$) 
However, as described below, more accurate LDA+U calculations estimate
$E_d \approx -1.6$ eV. Since the value $U=1.8$ eV was estimated \cite{Minamitani12_SU4_Kondo_in_FePc_molecules}, 
$E_d +U$ is small and in principle double occupancy of holes cannot be 
neglected contradicting the assumptions 
in Ref. \cite{Lobos14_SU4_Kondo_lattice}
(the hole occupancy is between 1 and 2). 
In addition, that theory does not allow to obtain the 
position of the split dips in the FKA at the same position as in the experiment, and an \emph{ad hoc}  
voltage (horizontal) shift in the differential conductance $dI/dV$ was necessary to fit the experiment. 
While a shift in $dI/dV$ (vertical) might be expected
due to uncertainties in the background contribution due to conduction electrons, a voltage shift is 
more difficult to justify. As a comparison,
for a Co atom on Au(111) a SBMFA of the SU(2) impurity Anderson model provides an excellent fit
of the observed $dI/dV$ without the need of a voltage shift \cite{aligia05}

The above discussion shows the need of accounting for a finite $U$ and 
multiple occupancy at the impurity sites. 
Kotliar and Ruckenstein originally implemented this idea within 
the SBMFA \cite{Kotliar86}, which was later generalized for 
orbitally-degenerate systems 
\cite{Dorin93_Slave_boson_finite_U, Hasegawa97_Slave_boson_finite_U,Fresard97_Slave_boson_finite_U}, 
and constitutes a very successful tool for describing 
highly-correlated systems 
\cite{dobro1997,merino2001,hardy2013}.
However, as we show below, this approach does not lead to the correct Kondo
temperature for an isolated molecule in the Kondo regime.

In this work we study the 
orbitally degenerate Hubbard-Anderson model 
as an
effective model to describe a square lattice of FePc molecules on Au(111),
for realistic parameters derived from the LDA+U. We construct a
SBMFA that gives the correct Kondo temperature for one FePc molecule,
and also reproduces very accurately the non-interacting limits (NILs) $U=0$, for both
the impurity and the lattice cases.  
Although physically we expect $U$ to be rather large (i.e., approx. 1.8 eV), 
reproducing the exactly solvable NIL constitutes an important ``sanity check'' 
for the entire formalism and is important when multiple occupancies cannot be neglected.
To the best of our knowledge, this is a novel theoretical development which improves over previous SBMFAs 
for orbitally-degenerate correlated systems 
\cite{Dorin93_Slave_boson_finite_U, Hasegawa97_Slave_boson_finite_U,Fresard97_Slave_boson_finite_U}. 
We also stress that the formalism presented here is in principle generic and applicable beyond 
the case of FePc/Au(111) molecules.

Our method allows for quantitatively accurate fits of the experimental $dI/dV$,  both for the
isolated molecule and for the square lattice. The fitting procedure allows to extract the parameters 
of our model. From here we conclude that while the single FePc molecule is in the Kondo limit, 
the lattice of FePc molecules is in the intermediate-valence regime.

The Hamiltonian was derived in Ref. \cite{Lobos14_SU4_Kondo_lattice}.
We refer the reader to that work for details. It can be split into
three parts. $H_{mol}$ is a degenerate Hubbard model that describes the
effective molecular states and the hopping between them, $H_{c}$ contains the
conduction states, and $H_{mix}$ is the coupling between molecular and
conduction states:

\begin{eqnarray}
H &=&H_{mol}+H_{c}+H_{mix},  \notag \\
H_{mol} &=&\sum\limits_{ij}^{N} [ -\sum\limits_{\sigma ,\nu }\left(%
t_{2}h_{\mathbf{r}_{ij},\sigma }^{\nu \dag }h_{\mathbf{r}_{ij}\pm \mathbf{a}%
_{\nu },\sigma }^{\nu }+t_{1}h_{\mathbf{r}_{ij},\sigma }^{\bar{\nu}\dag }h_{%
\mathbf{r}_{ij}\pm \mathbf{a}_{\nu },\sigma }^{\bar{\nu}}\right) \notag \\ 
&&+\ E_{d}n_{%
\mathbf{r}_{ij}}+\frac{U}{2}n_{\mathbf{r}_{ij}}\left( n_{\mathbf{r}%
_{ij}}-1\right) ] ,  \notag \\
H_{c} &=&\sum\limits_{ij}^{N}\sum_{\xi \sigma \nu }\epsilon _{\xi }c_{%
\mathbf{r}_{ij},\xi ,\sigma }^{\nu \dagger }c_{\mathbf{r}_{ij},\xi ,\sigma%
}^{\nu },\text{ } \notag \\ 
H_{mix} &=&V\sum\limits_{ij}^{N}\sum_{\xi \sigma \nu }\left(%
h_{\mathbf{r}_{ij},\sigma }^{\nu \dagger }c_{\mathbf{r}_{ij},\xi ,\sigma%
}^{\nu }+\mathrm{H.c.}\right). \label{h}%
\end{eqnarray}%
\newline
The operators $h_{\mathbf{r}_{ij},\sigma }^{\nu }$ annihilate a hole
(create an electron) in the state $|\nu _{\mathbf{r}_{ij},\sigma }\rangle $,
where $\nu =\left( x,y\right) $ denotes one of the two degenerate molecular
states with spin $\sigma $ at site with position $\mathbf{r}_{ij}=i\mathbf{a}%
_{1}+j\mathbf{a}_{2}$ of the square lattice, with $\mathbf{a}_{i}$ the
Bravais lattice vectors in the directions $x$ and $y$ respectively. For
simplicity we denote as $x$ and $y$ the orbitals with symmetry $xz$ and $yz$.
The operator for the
total number of holes at the molecule lying at site $\mathbf{r}_{ij}$ is $n_{\mathbf{r}%
_{ij}}=\sum_{\sigma \nu }n_{\mathbf{r}_{ij},\sigma }^{\nu }$, with $n_{%
\mathbf{r}_{ij},\sigma }^{\nu }=h_{\mathbf{r}_{ij},\sigma }^{\nu \dagger }h_{%
\mathbf{r}_{ij},\sigma }^{\nu }$. The meaning of $\bar{\nu}$ is $\bar{x}=y$,
and$\ \bar{y}=x$. The hopping $t_{2}$ between $x$ ($y$) orbitals in the $x$ (%
$y$) direction is larger than the hopping $t_{1}$ between $x$ ($y$) orbitals
in the $y$ ($x$) direction.

$H_{c}$ corresponds to a band of bulk and surface conduction electrons of
the substrate. The operator $c_{\mathbf{r}_{ij},\xi ,\sigma }$ annihilates a
conduction hole with spin $\sigma $ and quantum number $\xi $ at position $%
\mathbf{r}_{ij}$. Note that the form of $H_{mix}$ and $H_{c}$ assumes that
the molecular states of each Hubbard site $\mathbf{r}_{ij}$ is hybridized
with a different conduction band. This is an approximation which is valid if
the distance between the Hubbard sites is $R\gg 1/k_{F}$, with $k_{F}$ the
Fermi momentum of the metallic substrate \cite{Lobos12_Dissipative_XY_chain,
Lobos13_FMchains, Romero11_STM_for_adsorbed_molecules}, as it is the case in our problem.
In other words, the molecular states are well separated and if the 
hopping between them is neglected, the system behaves as dilute system of impurities.

We use the subscript $\eta $ to denote any of the four states at each site ($%
x\uparrow $, $x\downarrow $, $y\uparrow $, $y\downarrow $). The basic idea
of the SB approach is to enlarge the Fock space to include bosonic states
which correspond to each state in the fermionic description. For example,
the vacuum state at site $i$ is now represented as $e_{i}^{\dagger
}|0\rangle $, where $e_{i}^{\dagger }$ is a bosonic operator corresponding
to the empty site; similarly $s_{i\eta }^{\dagger }f_{i\eta }^{\dag
}|0\rangle $ represents the simply occupied state with one hole with ``color" 
$\eta $, $d_{i\eta _{1}\eta _{2}}^{\dagger }f_{i\eta _{1}}^{\dag }f_{i\eta
_{2}}^{\dag }|0\rangle $ corresponds to a state with double occupancy and
similarly for triple and quadruple hole occupancy (we denote the bosons as $%
t_{i\eta _{1}\eta _{2}\eta _{3}\text{ }}^{\dagger }$and $q$ respectively).
The fermion operators entering Eq. (\ref{h}) in the new representation are
given by

\begin{eqnarray}
h_{\mathbf{r}_{ij},\eta }^{\dag } &=&z_{\mathbf{r}_{ij},\eta }f_{\mathbf{r}_{ij},\eta }^{\dag },  \label{hole} \\
z_{\mathbf{r}_{ij},\eta }^{0} &=&s_{\mathbf{r}_{ij},\eta }^{\dagger }e_{\mathbf{r}_{ij}} +\sum_{\eta _{1}}d_{i\eta _{1}\eta
}^{\dag }s_{\mathbf{r}_{ij},\eta _{1}}\notag \\
 &+& \sum_{\eta _{1}<\eta _{2}}t_{i\eta _{1}\eta
_{2}\eta }^{\dag }d_{\mathbf{r}_{ij},\eta _{1}\eta _{2}}+q^\dagger t_{\mathbf{r}_{ij},\eta _{1}\eta _{2}\eta _{3}}, \label{z0}
\end{eqnarray}%
where all $\eta _{i}\neq \eta $ and where we assume $z_{\mathbf{r}_{ij},\eta }=z_{\mathbf{r}_{ij},\eta
}^{0}$ for the moment . The bosonic operator $z_{\mathbf{r}_{ij},\eta }^{0}$ corresponds to the creation
of the fermion $\eta $ at site $\mathbf{r}_{ij}$, in the bosonic sector of the Hilbert space.

To restrict the bosonic Hilbert space to the physical subspace, the
following constraints must be satisfied at each lattice site (in what follows we drop the
site index $\mathbf{r}_{ij}$ for simplicity)

\begin{eqnarray}
1 &=&e^{\dag }e+\sum_{\eta }s_{\eta }^{\dag }s_{\eta }+\sum_{\eta _{1}<\eta
_{2}}d_{\eta _{1}\eta _{2}}^{\dag }d_{\eta _{1}\eta _{2}} \notag \\ 
  &+& \sum_{\eta
_{1}<\eta _{2}<\eta _{3}}t_{\eta _{1}\eta _{2}\eta _{3}}^{\dag }t_{\eta
_{1}\eta _{2}\eta _{3}}+q^{\dag }q,  \notag \\
h_{\eta }^{\dag }h_{\eta } &=&s_{\eta }^{\dag }s_{\eta }+\sum_{\eta _{1}\neq
\eta }d_{\eta \eta _{1}}^{\dag }d_{\eta \eta _{1}}+\sum_{\eta _{1}<\eta
_{2}\neq \eta}t_{\eta \eta _{1}\eta _{2}}^{\dag }t_{\eta \eta
_{1}\eta _{2}} \notag \\ 
 &+& q^{\dag }q.  \label{cons}
\end{eqnarray}%
The interaction term in Eq. (\ref{h}) can be written as

\begin{eqnarray}
H_{U} &=& U( \sum_{\eta _{1}<\eta _{2}}d_{\eta _{1}\eta _{2}}^{\dag
}d_{\eta _{1}\eta _{2}}+3\sum_{\eta _{1}<\eta _{2}<\eta _{3}}t_{\eta
_{1}\eta _{2}\eta _{3}}^{\dag }t_{\eta _{1}\eta _{2}\eta _{3}} \notag \\
 &+& 6q^{\dag}q) .  \label{hu}
\end{eqnarray}%
The advantage of the SB representation is that using Eq. (\ref{hu}) in the
saddle-point approximation, in which all bosonic variables are replaced by
numbers, the problem is reduced to a non-interacting one, in which the
values of the bosons are obtained minimizing the free energy.
A shortcoming
of this procedure is that using Eqs. (\ref{hole},\ref{z0}), the exactly solvable 
NIL $U=0$ is not recovered after performing the mean-field approximation (MFA). In this
limit, all bosonic numbers become independent of position $\mathbf{r}_{ij}$ and color $\eta $, 
they can be chosen real and positive and for a total fermion 
occupation $n=\langle n_{\mathbf{r}_{ij}}\rangle $, calling $p=n/4$, one has 
$e^{2}=(1-p)^{4}$ (probability of finding an empty site), $s^{2}=p(1-p)^{3}$%
, $d^{2}=p^{2}(1-p)^{2}$, $t^{2}=p^{3}(1-p)$ and $q^{2}=p^{4}.$ This leads
to $z_{i\eta }^{0}=[p(1-p)]^{1/2}$, while the correct result is 
$z_{\eta}^{0}=1$ for $U=0$. To remedy this problem for the SU(2) case, Kotliar and
Ruckenstein have replaced the $z_{i\eta }^{0}$ operator for another one $%
z_{i\eta }$ which is equivalent to $z_{i\eta }^{0}$ in the restricted
Hilbert space, but when evaluated in the MFA for $U=0$ gives $z_{\eta }=1$ 
\cite{Kotliar86} They have shown that with this correction, the SBMFA is equivalent
to the Gutzwiller approximation. This procedure was generalized for an
arbitrary number $M$ of colors \cite{Dorin93_Slave_boson_finite_U,Hasegawa97_Slave_boson_finite_U,Fresard97_Slave_boson_finite_U}. For $M=4$ we can write
(dropping again the site indices for simplicity)

\begin{eqnarray}
z_{\eta } &=&O_{\eta L}^{-1/2}\ z_{\eta }^{0}\ O_{\eta R}^{-1/2},  \label{h2} \\
O_{\eta L} &=&1-A_{4}\ e^{\dag }e-A_{3}\sum_{\eta _{1}\neq \eta }s_{\eta
_{1}}^{\dag }s_{\eta _{1}} \notag \\
-&A_{2}&\sum_{\eta _{1}<\eta _{2}\neq \eta}d_{\eta _{1}\eta _{2}}^{\dag }d_{\eta _{1}\eta _{2}}-A_{1}\ t_{\eta
_{1}\eta _{2}\eta _{3}}^{\dag }t_{\eta _{1}\eta _{2}\eta _{3}},  \label{ol}
\\
O_{\eta R} &=&1-A_{1}\ s_{\eta }^{\dag }s_{\eta }-A_{2}\sum_{\eta _{1}\neq
\eta }d_{\eta \eta _{1}}^{\dag }d_{\eta \eta _{1}} \notag \\
-&A_{3}&\sum_{\eta _{1}<\eta
_{2}\neq \eta }t_{\eta \eta _{1}\eta _{2}}^{\dag }t_{\eta \eta
_{1}\eta _{2}}-A_{4}\ q^{\dag }q,  \label{or}
\end{eqnarray}%
where in Eq. (\ref{ol}) all $\eta _{i}\neq \eta $. \ We have written the
coefficients $A_{j}$ in Eqs. (\ref{ol},\ref{or}) so that 
the formalism is electron-hole symmetric, 
giving the same $z_\eta $ for occupation $n$ and $4-n$. Note that $z_{\eta }^{0}$ gives a
non vanishing result on states which do not contain $\eta $, while all terms
of Eq. (\ref{or}) except the first do contain this particle. Therefore $%
z_{\eta }^{0}O_{\eta R}^{-1/2}=z_{\eta }^{0}$. Similarly, since $z_{\eta
}^{0}$ creates this particle in the bosonic language, and all terms except
the first in Eq. (\ref{or}) project over states which do not contain it, $%
O_{\eta L}^{-1/2}z_{\eta }^{0}=z_{\eta }^{0}$. Then $z_{\eta }=z_{\eta }^{0}$
in an exact treatment. However this is no more true in the MFA. Choosing $%
A_{j}=1$, the second of Eqs. (\ref{cons}) gives $\langle O_{\eta R}\rangle =1-p$%
, and by electron-hole symmetry $\langle O_{\eta L}\rangle =p$. This implies 
$\langle z_{\eta }\rangle =1$ for $U=0$ recovering the correct
NIL in the SBMFA for any occupation.

Nevertheless, the widely used choice $A_{j}=1$ leads to an incorrect Kondo
temperature $T_{K}$ for the impurity case in the limit $U\rightarrow +\infty 
$. From exact Bethe ansatz results, the Kondo temperature in this limit for
the SU(N) Anderson model is (except for a factor of the order of 1 that
depends on the way in which it is defined) \cite{hewson}

\begin{equation}
T_{K}\simeq D\exp \left( \frac{\pi E_{d}}{N\Delta }\right) ,  \label{tk}
\end{equation}%
where a constant density of conduction states $\rho $ per color (spin and
symmetry) extending from $-D$ to $D$ and the Kondo regime $-E_{d}\gg \Delta $
are assumed, with $\Delta =\pi \rho V^{2}$. The problem is reduced to an effective one-fermion model, in
which the hybridization term is reduced by the factor $\langle z_{i\eta
}\rangle =z$, the term $H_{U}$ is added, and the constraints of Eqs. (\ref{cons}) should be added. 
Using the first constraint one can eliminate one of the
bosons, and the second one is taken into account by introducing a Lagrange
multiplier $\lambda $, leading to an effective impurity level at $\tilde{E}%
_{d}=E_{d}+\lambda $. The Green function of the pseudofermions is

\begin{equation}
G_{\eta }^{f}(\omega )=(\omega -\tilde{E}_{d}+i\tilde{\Delta})^{-1},
\label{gf}
\end{equation}%
where $\tilde{\Delta}=z^{2}\Delta $ is one of the definitions used for $%
T_{K}.$ The change in energy due to the impurity becomes \cite{hewson}

\begin{eqnarray}
E &=&\frac{1}{\pi }[-4\tilde{\Delta}+2\tilde{\Delta}\ln \left( \frac{\tilde{E%
}_{d}^{2}+\tilde{\Delta}^{2}}{D^{2}}\right) +4\tilde{E}_{d}\arctan \left( 
\frac{\tilde{\Delta}}{\tilde{E}_{d}}\right) ]  \notag \\
&&-4\lambda (s^{2}+3d^{2}+3t^{2}+q^{2})+6U(d^{2}+2t^{2}+q^{2}).  \label{e}
\end{eqnarray}

For $U\rightarrow +\infty $, one has $d=t=q=0$, and one can use $%
e^{2}=1-4s^{2}$ [from Eq. (\ref{cons})]. The condition $\partial E/\partial
\lambda =0$ gives $\tilde{E}_{d}=\tilde{\Delta}\cot (\pi s^{2})$ and
minimizing Eq. (\ref{e}) with respect to $s$ one obtains

\begin{equation}
\frac{2\Delta }{\pi }\frac{\partial z^{2}}{\partial s}\ln \left( \frac{%
\tilde{E}_{d}^{2}+\tilde{\Delta}^{2}}{D^{2}}\right) +8sE_{d}=0.  \label{mins}
\end{equation}%
In the Kondo limit, $-E_{d}\gg \Delta $, the total occupation $n\rightarrow
1 $, and then $s\rightarrow 1/2$ and Eq. (\ref{mins}) gives

\begin{equation}
\tilde{\Delta}=\frac{D}{\sqrt{2}}\exp \frac{\pi E_{d}}{C\Delta },\text{ }C=-%
\frac{\partial z^{2}}{\partial s}|_{s=1/2}.  \label{deltar}
\end{equation}%
In order for this equation to be consistent with Eq. (\ref{tk}), minus the
derivative of $z^{2}$ evaluated at $s=1/2$ should be $C=N=4$. Using Eqs. 
(\ref{z0},\ref{h2},\ref{ol}, and \ref{or}), and taking $A_{4}=1$ (to have a finite $z$ for 
$n\rightarrow 0)$ one obtains

\begin{eqnarray}
z^{2} &=&\frac{1-4s^{2}}{(4-3A_{3})(1-A_{1}s^{2})}, \ \ U\rightarrow +\infty ,
\label{zinf} \\
C &=&\text{ }\frac{16}{(4-3A_{3})(4-A_{1})}.  \label{c}
\end{eqnarray}%
Thus, using $A_{j}=1$ leads to an incorrect exponential dependence of $T_{K}$
on $E_{d}$ for $U\rightarrow +\infty $. To remedy this, we modify the
coefficients $A_{j}$ so that the condition $C=4$ is satisfied, and at the
same time we try to recover the NIL $U=0$ as accurately as
possible. To reproduce the NIL for $n=4p\rightarrow 0$, in
which $e^{2}\simeq 1-4p$, $s^{2}\simeq p$, and the rest of the bosons can be
neglected, we must choose $A_{4}=A_{3}=1$. This gives $O_{\eta R}=p$ [see
Eq. (\ref{or})] and compensates the factor $\sqrt{p}$ of $z_{0}$, while $%
O_{\eta L}\rightarrow 1$. Then, the condition $C=4$ implies $A_{1}=0$. To
determine the remaining coefficient we impose that the NIL
for $n=1$ is also recovered. This leads to a quadratic equation for $A_{2}$.
Its solution nearest to 1 gives $A_{2}=1.3185$. Note that because of
electron-hole symmetry, $z=1$ also for $n=3$ and $n\rightarrow 4$. We have represented
the resulting $z$ as a function of $n$ for $U=0$. It is a very flat function
with a value near 1, and the maximum deviation is at $n=2$ for which $z=0.989
$. Therefore, our SBMFA reproduces accurately not only the Kondo limit for
one impurity at large $U$, but also the NIL in the general case.
Another possibility to optimize the $A_j$ is to relax 
the correct NIL for $n \rightarrow 0$ and $n \rightarrow 4$ and to ask
that the formalism captures the correct Kondo temperature 
$T_{K}\simeq D\exp[-\pi U/(16 \Delta)]$ for $E_d=-3U/2$ such that 
the system is in the electron-hole symmetric case
(with $n=2$) for large but finite $U$. This condition would imply $A_2=2 \pm \sqrt{3/2}$. 
However, for realistic parameters
(as described below) this estimate gives $T_{K}\simeq 5 \times 10^{-7}$ K which is too small,
indicating that these values of $E_d$ are not realistic. Thus, we keep the values 
of $A_j$ described above.

The differential conductance $G(V_{b})=dI/dV_{b}$ measured in STS
experiments, is a non-equilibrium process in which the conduction states of
the tip of the scanning-tunneling microscope at a bias voltage $V_{b}$
hybridizes with a linear combination $l_{\mathbf{r}_{ij}}$ of molecular and conduction
states. This amounts to adding a perturbation $V_t \sum_k (t_k^{\dagger}l_{\mathbf{r}_{ij}} +\mathrm{H.c.})$ 
to the Hamiltonian, where $t_k$ refers to the tip states and 
$l_{\mathbf{r}_{ij}}=\alpha\left[\sum_{\eta }(c_{\mathbf{r}_{ij},\eta}+qh_{\mathbf{r}_{ij},\eta})\right]$ where $\alpha$ 
is a normalization constant.
In the limit in which $V_t$ is small, $G(V_{b})$ is
proportional to the density of states $\rho _{l}(eV_{b})$ of the states 
$l_{\mathbf{r}_{ij}}$ at the position $\mathbf{r}_{ij}$ of
the tip \cite{aligia05,meir92}. Here $q$ controls the ratio of the hybridization
of the tip with the molecular states with respect to the corresponding value
for the conduction states 
(this interference is the physical origin of the Fano effect). 
Once the Green function of the molecular states 
$G_{\mathbf{r}_{ij}\eta }^{h}(\omega )=z_{\mathbf{r}_{ij}, \eta }^{2}G_{\mathbf{r}_{ij},\eta }^{f}(\omega )$ 
[see Eq. (\ref{gf}) for the impurity case] is obtained from the SBMFA, 
$\rho _{l}(\omega)=-\mathrm{Im}[G_{l}(\omega )]/\pi $ is given by a simple expression obtained
using equations of motion \cite{aligia05}. One has 

\begin{equation}
G_{\mathbf{r}_{ij}}^{l}(\omega )=\sum _{\eta }[G_{\mathbf{r}_{ij},\eta }^{0c}(\omega )+(VG_{\mathbf{r}_{ij}, \eta
}^{0c}(\omega )+q)^{2}G_{\mathbf{r}_{ij}, \eta }^{f}(\omega )],  \label{rho}
\end{equation}%
where $G_{\mathbf{r}_{ij}, \eta }^{0c}$ is the Green function of the conduction states
before adding the molecules (we take that corresponding to a constant
density of states $\rho =1/(2D)$ extending from $-D$ to $D$). 

According to the estimations of LDA+U (see supplemental material of Ref. \cite{Minamitani12_SU4_Kondo_in_FePc_molecules}), 
for an isolated FePc molecule on Au(111), $U=1.84$ eV and the energy of the localized 
molecular orbital is $E_0= -3.91$ eV. Therefore, the energy necessary to add the 
first hole on a molecular orbital is $E_d=-(E_0+3U)=-1.61$ eV. In the rest of this work we take 
$U=1.8$ eV and for this system we take $E_d=-1.6$ eV (for the square lattice of FePc
molecules we modify $E_d$ as described below). We have chosen $D=3.65$ eV and $\rho =1/(2D)$.
However, the relevant parameter which controls the hybridization is the product 
$\Delta=\pi \rho V^2$, while the spectral density of the molecular states, as well as 
$\rho _{l}(\omega)/\rho_0$ depend very weakly on $D$.

\begin{figure}[h]
\includegraphics[width=0.9\linewidth,clip=true]{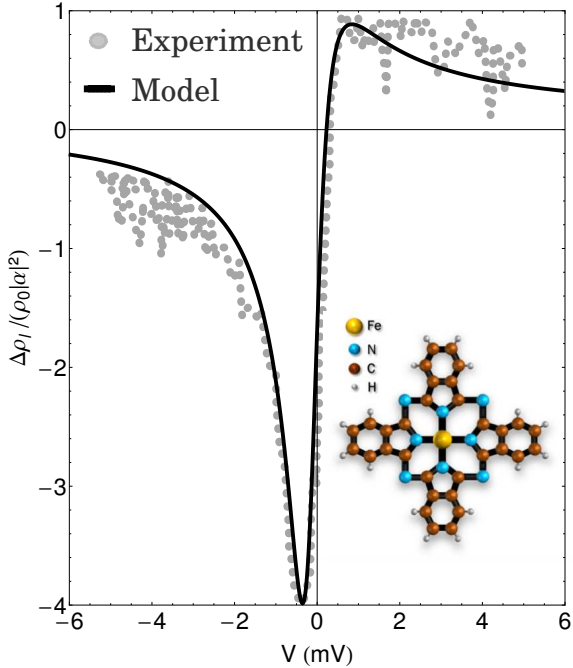}
\caption{change of the spectral density of the operator sensed by the STM tip $l$ 
(see text) as a function of frequency for an isolated FePc molecule. The circles 
correspond to $3.1G(\omega/eV)-2.5$ where $G$ in arbitrary units was taken from Fig. 3(a)
of Ref. \cite{Minamitani12_SU4_Kondo_in_FePc_molecules}. Parameters are $U=1.8$ eV, $E_d=-1.6$ eV, 
$\Delta=11$ meV and $q=-0.03$. The inset shows a scheme of the FePc molecule.}
\label{imp}
\end{figure}

To describe the case of a single FePc molecule, we 
take $\Delta$ and $q$ as fitting parameters. The resulting $\rho _{l}(\omega)$, except 
for an additive constant, is proportional to the observed $G$.
In Fig. \ref{imp} we represent the change $\Delta \rho _{l}(\omega)$ in $\rho _{l}(\omega)$ after addition of the
impurity [last term in Eq. (\ref{rho})] 
which should be proportional to the change in $G$, and is less sensitive to the details 
of the conduction band. 
From the fit we obtain $\Delta=11$ meV and $q=-0.03$.
Note that the agreement with experiment near the Fermi energy is very good.  
It is certainly better than the comparison between two different experimental
realizations [Figs. 3 (a) and (b) of Ref. \cite{Minamitani12_SU4_Kondo_in_FePc_molecules} for zero magnetic field].
For the parameters of our fit, we obtain a total occupation of 1.01 holes (2.99 electrons) in the molecular states. The total probability of double hole occupancy is 1.3\% 
and that of having no holes (4 electrons) is 0.14\%. Therefore,  we conclude that the single FePc molecule 
is in the Kondo regime.
These results agree with those of Minamitani \etal \cite{Minamitani12_SU4_Kondo_in_FePc_molecules} who used the  
numerical-renormalization group to interpret the experiments. However, those authors
did not attempt to fit the experiment.

To model the square lattice of FePc molecules on Au(111), we need to introduce 
the intermolecular hopping elements $t_i$. We assume $t_2/t_1=3$ as estimated previously \cite{Lobos14_SU4_Kondo_lattice}.
The SBMFA again reduces the problem to an effective non-interacting one. The latter has the same 
form as that explained in the supplemental material of Ref. \cite{Lobos14_SU4_Kondo_lattice}, but 
in our case a finite $U$ is used and therefore the quasiparticle weight $z_{\mathbf{r}_{ij},\eta }^2$ is
different and given by Eqs. (\ref{z0},\ref{h2},\ref{ol}, and \ref{or}) 
with $A_1=0$, $A_{2}=1.3185$, and $A_3=A_4=1$ 
as described above. Although for the lattice, the 
model is not SU(4)
symmetric, the effective non-interacting problem retains this symmetry \cite{sym} and in the homogeneous 
SU(4) symmetric solution $z_{\mathbf{r}_{ij},\eta }=z$ independent of site, spin and orbital.
In order to compare with the experimental results for the square lattice \cite{Tsukahara10_Evolution_of_Kondo_resonance},
we retain the same values of $U$ and $\Delta$ as for the impurity case, and use $t_1$, $q$ and $E_d$ as 
fitting parameters. $q$ depends on the distance from the STM tip to molecules and the surface
and therefore one expects that it is different from the one-molecule case. Concerning 
$E_d$ since the molecules are negatively charged as they are placed on the surface \cite{Minamitani12_SU4_Kondo_in_FePc_molecules,Minamitani12_Ab_initio},
one expects that $E_d$ lowers for the lattice due to interatomic repulsion, since holes are stabilized.
This agrees qualitatively with our findings. The quantitative aspects will be discussed below.

\begin{figure}[h]
\includegraphics[width=0.85\linewidth,clip=true]{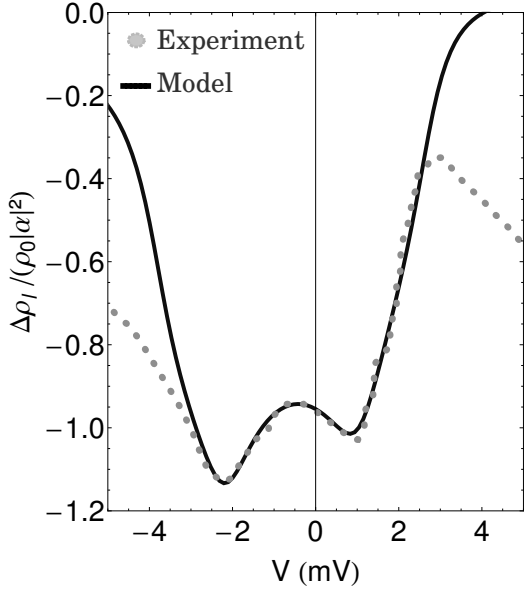}
\caption{
same as Fig. 1 for a square lattice of FePc molecules. The circles 
correspond to $0.28G(\omega/eV)-0.7$ where G in arbitrary units was taken from
Ref. \cite{Tsukahara10_Evolution_of_Kondo_resonance}.
Parameters are $U=1.8$ eV, $E_d=-1.828$ eV, $\Delta=11$ meV $t_1=7.5$ meV,  and $q=-0.006$.}
\label{red}
\end{figure}

The resulting change in the spectral density sensed by the STM tip is shown in Fig. \ref{red}.
From the fit we obtain $t_1=7.5$ meV, $q=-0.006$ and $E_d=-1.828$ eV. The latter is a surprising result
since $E_d+U=-28$ meV has become slightly negative, favoring double occupancy but the system is in the 
intermediate valence regime, because this energy is of the order of $\Delta=11$ meV. This value of
$E_d$ is imposed by the position in voltage of the observed differential conductance. 
Higher values of $E_d$ increase the electron occupation and shift the structure to the left.
For the parameters of Fig. \ref{red} we obtain a total occupancy of 1.65 holes (2.35 electrons).
The total single hole occupancy is 0.34, the total probability of double hole occupancy is 0.65,
and other states can be neglected.

The drastic change in the charge of the molecular $xz$ and $yz$ orbitals is surprising.
We should note that first-principle studies of the charges for isolated FePc \cite{Minamitani12_SU4_Kondo_in_FePc_molecules,Minamitani12_Ab_initio} and
CoPc \cite{Wang14} on Au(111) indicate that in addition to these orbitals, also the $3d$ orbitals of the TMAs
with symmetry $3z^2-r^2$ are partially occupied. Nevertheless, Minamitani \etal \ showed that the spin
of these orbitals is screened in a first-stage Kondo effect at higher energy and that an effective model 
containing only $xz$ and $yz$ orbitals as the localized ones describes the low-energy physics \cite{Minamitani12_SU4_Kondo_in_FePc_molecules}.
When an isolated FePc molecule is placed on the metal surface, a partial transfer of electrons takes 
place from the metallic states and also from the $xz$ and $yz$ orbitals to the $3z^2-r^2$ \cite{Minamitani12_Ab_initio}.
Our results suggest that the latter charge transfer is enhanced for a lattice of FePc molecules on Au(111).
The intra-atomic inter-orbital repulsion should favor this procedure, since a larger charge in the 
$3z^2-r^2$ orbital increases the energy for occupying $xz$ and $yz$ orbitals. These arguments suggest
that a more general model that includes $3z^2-r^2$ localized states 
might explain the change in the occupancy of  $xz$ and $yz$ orbitals.

In summary, we have studied an array of FePc molecules deposited on top of Au(111), described by an effective Hubbard-Anderson model with degenerate effective orbitals with $xz$ and $yz$ symmetry.
To that end, we have introduced a generalized slave-boson mean-field approximation (SBMFA), which correctly
describes both the non-interacting ($U=0$) and the strongly-interacting ($U \rightarrow \infty$) limits. 
This is an important improvement over previous formulations of the SBMFA for strongly-correlated 
orbitally-degenerate systems, which fail to describe the  Kondo limit. We stress that our method is generic 
and has applications beyond the present case of FePc/Au(111). 
Our results indicate that while the isolated FePc molecule can be described by an SU(4) Anderson model 
in the Kondo regime, the case of the square lattice corresponds to the intermediate-valence regime, 
with a total occupation of nearly 1.65 holes
in the FePc molecular orbitals. 
This conclusion is imposed by the position in voltage of the double-dip structure
observed in $dI/dV$ and is independent of the details of the parameters.
We believe that the shift in $E_{d}$ is partly due to
intermolecular repulsion, but it might also indicate a redistribution of charge
among the Fe $3d$ electrons that is beyond our effective model.

We thank G. Chiappe for useful discussions on the electronic structure of
the isolated FePc molecule. JF and AAA are partially supported by CONICET,
Argentina. This work was sponsored by PICT 2010-1060 and 2013-1045 of the
ANPCyT-Argentina and PIP 112-201101-00832 of CONICET. 
AML acknowledges support from NSF-PFC-JQI.


\begin{thebibliography}{10}
\expandafter\ifx\csname url\endcsname\relax\def\url#1{\texttt{#1}}\fi

\bibitem{Wolf01_Spintronics_review}
\Name{Wolf S.~A. \textit{et al}}
  \REVIEW{Science}{294}{2001}{5546}.

\bibitem{Li98_Kondo_effect_on_single_adatoms}
\Name{Li J., Schneider W.-D., Berndt R. \and Delley B.} \REVIEW{Phys. Rev.
  Lett.}{80}{1998}{2893}.

\bibitem{Madhavan98_Tunneling_into_single_Kondo_adatom}
\Name{Madhavan V., Chen W., Jamneala T., Crommie M.~F. \and Wingreen N.~S.}
  \REVIEW{Science}{280}{1998}{567}.

\bibitem{knorr02}
\Name{Knorr N., Schneider M.~A., Diekhoner L., Wahl P. \and Kern K.}
  \REVIEW{Phys. Rev. Lett.}{88}{2002}{096804}.

\bibitem{Zhao05_Science}
\Name{Zhao A. \textit{et al}} \REVIEW{Science}{309}{2005}{1542}.

\bibitem{Iancu06_Manipulation_of_Kondo_effect}
\Name{Iancu V., Deshpande A. \and Hla S.-W.} \REVIEW{Phys. Rev.
  Lett.}{97}{2006}{266603}.

\bibitem{Gao07_Site_specific_Kondo_effect}
\Name{Gao L. \textit{et al}} \REVIEW{Phys.
  Rev. Lett.}{99}{2007}{106402}.

\bibitem{Tsukahara10_Evolution_of_Kondo_resonance}
\Name{Tsukahara N.  \textit{et al}}
  \REVIEW{Phys. Rev. Lett.}{106}{2011}{187201}.

\bibitem{Franke11_Competition_of_Kondo_and_SC_in_molecules}
\Name{Franke K.~J., Schulze G. \and Pascual J.~I.}
  \REVIEW{Science}{332}{2011}{940}.

\bibitem{Minamitani12_SU4_Kondo_in_FePc_molecules}
\Name{Minamitani E.  \textit{et al}} \REVIEW{Phys. Rev. Lett.}{109}{2012}{086602}.

\bibitem{DiLullo12_Molecular_Kondo_chain}
\Name{DiLullo A.  \textit{et al}}
  \REVIEW{Nano Lett.}{12}{2012}{3174}.

\bibitem{parksjj}
\Name{Parks J.~J. \textit{et al}}
  \REVIEW{Science}{328}{2010}{1370}.

\bibitem{serge}
\Name{Florens S. \textit{et al}} \REVIEW{J. Phys. Cond. Matt.}{23}{2011}{243202}.

\bibitem{jarillo05}
\Name{Jarillo-Herrero P. \textit{et al}} \REVIEW{Nature (London)}{434}{2005}{484}.

\bibitem{tetta12}
\Name{Tettamanzi G.~C. \textit{et al}} \REVIEW{Phys. Rev. Lett.}{108}{2012}{046803}.

\bibitem{Lobos14_SU4_Kondo_lattice}
\Name{Lobos A.~M., Romero M. \and Aligia A.~A.} \REVIEW{Phys. Rev.
  B}{89}{2014}{121406}.

\bibitem{coleman84}
\Name{Coleman P.} \REVIEW{Phys. Rev. B}{29}{1984}{3035}.

\bibitem{aligia05}
\Name{Aligia A.~A. \and Lobos A.~M.} \REVIEW{J. Phys.: Condens.
  Matter}{17}{2005}{S1095}.

\bibitem{Kotliar86}
\Name{Kotliar G. \and Ruckenstein A.~E.} \REVIEW{Phys. Rev.
  Lett.}{57}{1986}{1362}.

\bibitem{Dorin93_Slave_boson_finite_U}
\Name{Dorin V. \and Schlottmann P.} \REVIEW{Phys. Rev. B}{47}{1993}{5095}.

\bibitem{Hasegawa97_Slave_boson_finite_U}
\Name{Hasegawa H.} \REVIEW{J. Phys. Soc. Jpn.}{66}{1997}{1391}.

\bibitem{Fresard97_Slave_boson_finite_U}
\Name{Fr\'{e}sard R. \and Kotliar G.} \REVIEW{Phys. Rev. B}{56}{1997}{12909}.

\bibitem{dobro1997}
\Name{Dobrosavljevi\ifmmode~\acute{c}\else \'{c}\fi{} V. \and Kotliar G.}
  \REVIEW{Phys. Rev. Lett.}{78}{1997}{3943}.

\bibitem{merino2001}
\Name{Merino J. \and McKenzie R.~H.} \REVIEW{Phys. Rev.
  Lett.}{87}{2001}{237002}.

\bibitem{hardy2013}
\Name{Hardy F. \textit{et al}}  \REVIEW{Phys. Rev. Lett.}{111}{2013}{027002}.

\bibitem{Lobos12_Dissipative_XY_chain}
\Name{Lobos A.~M., Cazalilla M.~A. \and Chudzinski P.} \REVIEW{Phys. Rev.
  B}{86}{2012}{035455}.

\bibitem{Lobos13_FMchains}
\Name{Lobos A.~M. \and Cazalilla M.~A.} \REVIEW{J. Phys.: Condens.
  Matter}{25}{2013}{094008}.

\bibitem{Romero11_STM_for_adsorbed_molecules}
\Name{Romero M. \and Aligia A.~A.} \REVIEW{Phys. Rev. B}{83}{2011}{155423}.

\bibitem{hewson}
\Name{Hewson A.~C.} \Book{The Kondo Problem to Heavy Fermions} (Cambridge
  University Press) 1993, chapter 7

\bibitem{meir92}
\Name{Meir Y. \and Wingreen N.~S.} \REVIEW{Phys. Rev. Lett.}{68}{1992}{2512}.

\bibitem{sym}
For $U=0$, $H$ commutes with the following non-trivial SU(4) generators: $S_{\nu
  \sigma }^{\nu \bar{\sigma}}=\sum_{ij}^{N}h_{\mathbf{r}_{ij},\sigma }^{\nu
  \dagger }h_{\mathbf{r}_{ij},,\bar{\sigma}}^{\nu }$ and $S_{\nu \sigma
  }^{\bar{\nu}\sigma ^{\prime }}=\sum_{ij}^{N}h_{\mathbf{r}_{ij},\sigma }^{\nu
  \dagger }h_{R\mathbf{r}_{ij},,\sigma ^{\prime }}^{\bar{\nu}}$, where $R$ is
  the reflection that permutes $x$ and $y$.

\bibitem{Minamitani12_Ab_initio}
\Name{Minamitani E. \textit{et al}} \REVIEW{e-J. Surf. Sci. Nanotech.}{10}{2012}{38 44}.

\bibitem{Wang14}
\Name{Wang Y., Zheng X., Li B. \and Yang J.} \REVIEW{J. Chem.
  Phys}{141}{2014}{084713}.

\end{thebibliography}
\end{document}